\documentclass[superscriptaddress,secnumarabic,amssymb,amsmath,nobibnotes,aps,prd,showkeys,showpacs,nofootinbib,preprint]{revtex4}

\def\[{\left\lbrack}
\def\]{\right\rbrack}

\def\({\left(}
\def\){\right)}
\def\ni{\noindent}

\newcommand{\be}{\begin{equation}}
\newcommand{\ee}{\end{equation}}
\newcommand{\ea}{\end{eqnarray}}
\newcommand{\ba}{\begin{eqnarray}}

\begin{document}

\title{\Large Helicity and Vortex generation}

\author{Albert C. R. Mendes}\email{albert@fisica.ufjf.br}
\affiliation{Departamento de F\'{i}sica, Universidade Federal de Juiz de Fora, 36036-330, Juiz de Fora, MG, Brazil}

\author{Flavio Takakura}\email{takakura@fisica.ufjf.br}
\affiliation{Departamento de F\'{i}sica, Universidade Federal de Juiz de Fora, 36036-330, Juiz de Fora, MG, Brazil}

\author{Everton M. C. Abreu}\email{evertonabreu@ufrrj.br}
\affiliation{Grupo de F\' isica Te\'orica e F\' isica Matem\'atica, Departamento de F\'{i}sica, Universidade Federal Rural do Rio de Janeiro, 23890-971, Serop\'edica, RJ, Brazil}
\affiliation{Departamento de F\'{i}sica, Universidade Federal de Juiz de Fora, 36036-330, Juiz de Fora, MG, Brazil}

\author{Patrick Paolo Silva}\email{patrick@fisica.ufjf.br}
\affiliation{Departamento de F\'{i}sica, Universidade Federal de Juiz de Fora, 36036-330, Juiz de Fora, MG, Brazil}

\author{Jo\~ao Vitor Frossad}\email{jvitor@fisica.ufjf.br}
\affiliation{Departamento de F\'{i}sica, Universidade Federal de Juiz de Fora, 36036-330, Juiz de Fora, MG, Brazil}

\author{Jorge Ananias Neto}\email{jorge@fisica.ufjf.br}
\affiliation{Departamento de F\'{i}sica, Universidade Federal de Juiz de Fora, 36036-330, Juiz de Fora, MG, Brazil}


\date{\today}

\begin{abstract}
\ni In this paper we have continued the calculations made recently concerning he generalization of the minimal coupling prescription.   We have obtained the Navier-Stokes equation for the charged fluid embedded into an electromagnetic field.   We have analyzed the wave feature of the fields.   Wave equations for velocity and vorticity were obtained and gauge choices were discussed.  In this paper we have studied the evolution of helicity and circulation from the Maxwell type formulation to the equations of a compressive fluid, charged in interaction with an electromagnetic field. We see that for both helicity and circulation there are terms that, in principle, can be considered as source terms or creation of circulation in the dynamics of fluids
\end{abstract}



\maketitle

\section{Introduction}
In recent papers some of us have discussed that, as an alternative way for the description of fluid dynamics, concerning both the compressible fluids \cite{Kambe} and the plasma equations \cite{Thompson}, the better path would be through the recasting of the equations of motion to obtain a set of Maxwell-type equations for the fluid. This transformation in the structure of the equations of motion results in the generalization of the concept of charge and current connected to the dynamics of the fluid \cite{Marmanis,jnpp}.  The identification of what will be considered as a source term in the resulting theory depends on the choice of the objects which will form the main part of its new structure of fluid dynamics. In Lighthill's work concerning the sound radiated by a fluid flow \cite{lighthill}, the applied stress tensor was considered as the source of the radiation field. R. J. Thompson \cite{Thompson} recently introduced an extension of this new structure of the plasma equations of motion, for each kind of fluid, from the equations of motion that  describe such system. 

The main motivation is to understand the thermodynamical arguments in order to obtain how the energy density $\rho$ depends on the temperature $T$ for a fluid's equation of state given by $p=\omega \rho$.  Besides, the Stefan-Boltzmann law has been widely discussed in the scenario of black holes thermodynamics 
\cite{hawking}, from where we know that the energy density is inversely proportional to the temperature.  More recently, the observed acceleration of the Universe demands the existence of a new component, the well known dark energy, which rules out all other forms of energy and has a negative  pressure.  The presence of such energy in the Universe deserves detailed analysis, such as the consequences related to the application of the generalized second law \cite{beckenstein} or the entropy bound \cite{beckenstein2}.   Some elements, such as the phantom fields ($\omega < -1$), which have a negative kinetic energy, negative temperature and the entropy is always positive, can also change completely the evolution of black holes and their connection to the generalized second law, as was discussed in \cite{fh,gs}.

The purpose of this paper is to analyze the equation that governs the evolution of helicity and also to study the mechanism of vortex generation from the rate of variation of the circulation associated with the canonical moment $\oint_{\gamma} {\vec v}\cdot d^3 x$. The origin of the magnetic field as well as the vorticity is one of the problems not solved in theoretical physics.

In this paper we will follow a sequence such that in section II, we have continued the calculations that began in \cite{albert1} concerning he generalization of \cite{mahajan}.   We have obtained the Navier-Stokes equation for the charged fluid embedded into an electromagnetic field.   In section III we have analyzed the wave feature of the fields.   Wave equations for velocity and vorticity were obtained and gauge choices were discussed.   In section IV, we have computed the dynamics of helicity and circulation.   We have left the final section for the general considerations and conclusions.

\section{Lagrangian Formalism and the field equations }

Recently \cite{Albert1,albert1,albert2,albert3}, some of us have shown that a Lagrangian formulation for a compressible charged fluid, with each species labeled by an index $\alpha$, immersed  in an electromagnetic field, can be obtained resulting in a Maxwell-type action for the fluid, given by 
\be\label{01}
{\cal L} = -{1\over 4} G_{\alpha}^{\mu\nu}G^{\alpha}_{\mu\nu} -J_{\alpha}^{\mu}U^{\alpha}_{\mu},
\ee
where $G_{\alpha}^{\mu\nu}$ is a second rank tensor taking into account the coupling with the electromagnetic field, given by,
\be\label{02}
G_{\alpha}^{\mu\nu}= T_{\alpha}^{\mu\nu} + {\epsilon_{\alpha}\over m_{\alpha}}F^{\mu\nu}\,\,,
\ee
where $\epsilon_\alpha$ is the charge and $m_\alpha$ is the mass of the charge. The fully antisymmetric second-rank ``flow'' tensor is defined by
\be\label{03}
T_{\alpha}^{\mu\nu} = \partial^{\mu} U_{\alpha}^{\nu} -\partial^{\nu} U_{\alpha}^{\mu},
\ee
and the four-vector potential
\be\label{04}
U_{\alpha}^{\mu}=(\phi_{\alpha}, {\vec v}_{\alpha})\,\,,
\ee
where $\phi_{\alpha} = h_{\alpha} +{1\over2}v^2_{\alpha}$ is the total energy, $h$ is the enthalpy per mass unit and ${\vec v}_{\alpha}$ is the average velocity field. The space-time metric is $\eta_{\mu\nu} =(-+++)$ and $F^{\mu\nu}$ is the second-rank tensor of the electromagnetic field.

So, from the Lagrangian density in Eq. (\ref{01}) we have the following set of inhomogeneous equations
\be\label{05}
\partial_{\mu} G_{\alpha}^{\mu\nu} = J_{\alpha}^{\nu}\,\,,
\ee
\be\label{06}
\nabla \cdot {\vec l}_{\alpha} = n_{\alpha} - {{\epsilon_{\alpha}}\over{m_{\alpha}}}{1\over\epsilon_0}\rho_{el} -\nabla \cdot{\vec k}_{\alpha}\,\,,
\ee
\be\label{07}
\nabla \times {\vec\omega}_{\alpha} -{{\partial {\vec l}_{\alpha}}\over{\partial t}}
={\vec j}_{\alpha} - {{\epsilon_{\alpha}}\over{m_{\alpha}}}{1\over\epsilon_0}{\vec j}_{el} + {{\partial{\vec k}_{\alpha}}\over{\partial t}}\,\,,
\ee
where $\vec \omega =\nabla \times \vec v$ is the vortex and $\vec l= \vec\omega \times\vec v$ is the Lamb vector, that can be written as
\be\label{07.1}
\vec l= -{{\partial \vec v}\over{\partial t}} -\nabla\phi\,\,.
\ee
The homogeneous equations that can be written in terms of the dual field-strength tensor are
\be\label{08}
\partial_\mu {\cal G}_{\alpha}^{\mu\nu}=0\,\,; \,\,\,\,\qquad\qquad {\cal G}^{\mu\nu} ={1\over 2}\epsilon^{\mu\nu\gamma\beta}G_{\gamma\beta}\,\,,
\ee
\be\label{09}
\nabla \cdot {\vec \omega}_{\alpha} =0\,\,,
\ee
\be\label{10}
\nabla \times {\vec l}_{\alpha} +{{\partial {\vec \omega}_{\alpha}}\over{\partial t}} =-\nabla\times{\vec k}_{\alpha}\,\,,
\ee
where ${\vec k}_{\alpha} =T\nabla s_{\alpha} +\rho_{\alpha}^{-1} \nabla \sigma$ that carries the contribution due to viscosity and statistical features; $T$ is the temperature, $s_{\alpha}$ is the entropy per unit mass, and 
\be\label{11}
\sigma_{ij} =\mu e_{ij} + \xi\delta_{ij} D
\ee
is the viscosity stress ($\mu$ and $\xi$ are called coefficients of viscosity) \cite{Landau}. 

The derivation of Eqs. (\ref{06}) and (\ref{07}) has demanded the introduction of two new concepts analogous to the electric charge densities $\rho_{el}$ and current densities ${\vec j}_{el}$, the turbulent charge $n_{\alpha}$ and the turbulent current vector ${\vec j}_{\alpha}$ given by
\be\label{12}
n_{\alpha} =-{\partial\over{\partial t}}\nabla \cdot{\vec v}_{\alpha} -\nabla^2 \phi_{\alpha},
\ee
\be\label{13}
{\vec j}_\alpha = \left(\hat{N}_\alpha + {{\epsilon_{\alpha}}\over{m_{\alpha}}}{{1}\over{\epsilon_{0}}} \rho_{el}+\nabla \cdot {\vec k}_\alpha\right)\hat{\vec v}_\alpha + 2\hat{\vec l}_\alpha \cdot \nabla \hat{\vec v}_\alpha -\hat{\vec l}_\alpha (\nabla \cdot \hat{\vec v}_\alpha) + \hat{\vec l}_\alpha \times \hat{\vec\omega}_\alpha - \hat{\vec\omega}_\alpha \times {{\partial \hat{\vec v}_\alpha}\over{\partial t}}\,\,.
\ee
The analytical form of the source terms describe the behavior of these effective charge and current densities as a function of the fluid fields. These quantities should be taken as an input and can not be determined by the theory.   Thus, our ability to obtain them by observation have an utterly importance and the construction of appropriate models for both the generalized charge and current densities is a necessary step to calculate the generalized fields of the fluid.

%
%
%
%
%
%

We can also write the above Lagrangian density Eq. (\ref{01}) in terms of the ``potentials of the theory," as
\ba\label{17}
{\cal L} &=& -{1\over 2}\left(\vec l_{\alpha} +{\epsilon_{\alpha}\over m_{\alpha}}\vec E \right)^2 -{1\over 2}\left(\vec \omega_{\alpha} + {\epsilon_{\alpha}\over m_{\alpha}}\vec B \right)^2 \nonumber\\
&=&{1\over 2}\left(-{{\partial {\vec v}_{\alpha}}\over{\partial t}}-\nabla \phi_{\alpha} + \vec k_{\alpha}+{\epsilon_{\alpha}\over m_{\alpha}}\vec E  \right)^2 -{1\over 2}\left(\nabla \times \vec v_{\alpha}+ {\epsilon_{\alpha}\over m_{\alpha}}\vec B \right)^2\,\,.
\ea
Notice that the interaction with the electromagnetic field let us to introduced a generalized Lamb vector ($\hat{\vec l}$) and generalized vortex ($\hat{\vec\omega}$) field 
\be\label{18}
\hat{\vec{l}}_{\alpha}= {\vec l}_{\alpha} +{\epsilon_{\alpha}\over m_{\alpha}}\vec E \,\,,\,\,\,\,\,\,\,\,\qquad
\hat{\vec{\omega}}_{\alpha}= \vec{\omega}_{\alpha} + {{\epsilon_{\alpha}}\over m_{\alpha}}\vec B\,\,.
\ee

Another  interesting consequences of this formalism, Eq. (\ref{01}), for the compressible fluid is given by the conjugated momenta associated with velocity, that is
\be\label{19}
\vec{\pi}_{\alpha} ={\delta {\cal L}\over{\delta\dot{\vec v}_{\alpha}}}={{\partial {\vec v}_{\alpha}}\over{\partial t}}+\nabla \phi_{\alpha} - \vec k_{\alpha}-{\epsilon_{\alpha}\over m_{\alpha}}\vec E =-\hat{\vec{l}}_{\alpha}
\ee
or 
\be\label{20}
{{\partial {\vec v}_{\alpha}}\over{\partial t}}+\nabla \phi_{\alpha} - \vec k_{\alpha}-{\epsilon_{\alpha}\over m_{\alpha}}\vec E= -\hat{\vec{\omega}}_{\alpha}\times \vec{v}_{\alpha}\,\,.
\ee
Using the relation (\ref{18}), we finally have the Navier-Stokes equation for the charged fluid immersed  in an electromagnetic field
\be\label{21}
{{\partial {\vec v}_{\alpha}}\over{\partial t}}+{\vec{\omega}}_{\alpha}\times \vec{v}_{\alpha}=-\nabla \phi_{\alpha} + {\epsilon_{\alpha}\over m_{\alpha}}\Big[\vec E+ \vec{v}_{\alpha} \times \vec B \Big] +\vec k_{\alpha}\,\,,
\ee
this is the momentum equation where the Lorentz force is present. This equation must be further supplemented by conservation of entropy and the Maxwell equations.

\section{Wave Equations} 

We can also analyze is the wave feature of the fields for the fluid immersed in the electromagnetic field. As in Maxwell's electromagnetism, under certain circumstances it is more appropriate to work with the wave equations  because they allow a better understanding of how the electromagnetic fields behave.   So, from the Eqs. (\ref{06}) and (\ref{07.1})
\be\label{22}
{{\partial^2 {\vec v}_{\alpha}}\over{\partial t^2}} -\nabla^2 {\vec v}_{\alpha}  = {\vec j}_{\alpha} + {{\epsilon_{\alpha}}\over{m_{\alpha}}}{\vec j}_{el} -\nabla\left( \nabla \cdot{\vec v}_{\alpha} +{{\partial \phi_{\alpha}}\over{\partial t}} \right) +{{\partial {\vec k}_{\alpha}}\over{\partial t}}\,\,,
\ee
\be\label{23}
{{\partial^2 {\vec \omega}_{\alpha}}\over{\partial t^2}} -\nabla^2 {\vec \omega}_{\alpha} =  \nabla \times \left( {\vec j}_{\alpha}  +{{\epsilon_{\alpha}}\over{m_{\alpha}}}{\vec j}_{el}\right) + \nabla\times{{\partial {\vec k}_{\alpha}}\over{\partial t}}\,\,,
\ee
which are the wave equations for the velocity and vorticity fields with the objects written in terms of turbulent sources and with contributions due to the interaction with the electromagnetic field, ${\vec j}_{el}$, as well as due to both viscosity and temperature in ${\vec k}_{\alpha}$.

Eq. (\ref{22}) shows the wave character of the fluid dynamics equations described by Eqs. (\ref{06})-(\ref{07}) and Eqs. (\ref{09})-(\ref{10}). Hence,  knowing the source terms, we may determine the evolution of each species within the fluid separately by these wave equations. Besides, Eqs. (\ref{22}) allow us to make an interesting observation about a conceptual difference between both theories, namely, electromagnetism and fluid theories.

The equations for the electromagnetic potentials  $(\vec A, \Phi)$, which have expressions analogous to mathematical equations, can be decoupled through a suitable choice for potential called gauge transformations
\be
\label{24}
\nabla \cdot \vec A =0 \,\,\,\,\, ({\rm Coulomb\,\, gauge})
\ee
and
\be\label{25}
\nabla \cdot \vec A + {{1}\over{c^2}} {{\partial \Phi}\over{\partial t}} = 0 \,\,\,\,\,\, ({\rm Lorentz\,\, gauge})\,\,,
\ee
which do not affect the physics of the system. In fluid dynamics, this freedom is not simply a choice of gauge, it has implications about the physical nature of the flow.  The relation $\nabla \cdot  \vec v =0$  in fluid dynamics, which is equivalent to the Coulomb gauge, is not a true gauge but it is a choice of the incompressibility of the flow.

Similarly, the Lorentz gauge has a corresponding equation which connects the fluid dynamics given by 
\be\label{26}
\nabla \cdot \vec{v} + {{\partial \phi }\over{\partial t}} =0,
\ee
and which is analogous to a compressible fluid. Thus, we observe that a ``gauge choice" in fluid dynamics is directly related to the hypothesis made about the compressibility or incompressibility of the fluid \cite{Thompson}.

\section{Evolution of the Helicity and Circulation} 

It is well known that for the electromagnetic field in vacuum, the helicity is defined by 
\be\label{27}
h=\int {\vec A}\cdot{\vec B}\,d^{3}x\,\,.
\ee
Here $\vec A$ is the usual magnetic vector potential and $\vec B=\nabla\times \vec A$. The relativistic generalization of helicity is defined by an integral over the zeroth component of the four-vector $j^{\nu}=A_{\mu}{\cal F}^{\mu\nu}$, $\int j^{0} d^3 x$, where ${\cal F}^{\mu\nu}$ is the dual of the second-rank tensor of the electromagnetic field $F^{\mu\nu}$. Its consequence is that $j^{\nu}$ is conserved 
\be\label{28}
\partial_\nu j^{\nu} =\,-\,2\vec E \cdot \vec B = 0
\ee
only if $\vec E \cdot \vec B = 0$. So the helicity $\int j^{0} d^3 x$ is a constant of the motion.

Therefore, from the analogy with electromagnetism, we can represent both ${\cal F}^{\mu\nu} \rightarrow {\cal G}^{\mu\nu}$ and $A_\mu \rightarrow  {\hat U}_\mu =U_\mu + {{\epsilon}\over {m}} A_\mu$ to the fluid, thus the four-vector $j^\mu$ can be written 
\be\label{29}
j^\mu = \left( U_\mu + {{\epsilon}\over {m}} A_\mu \right){\cal G}^{\mu\nu}\,\,.
\ee
Then, using the Eq. (\ref{09}), the antisymmetry of ${\cal G}^{\mu\nu}$, and
\be\label{30}
G_{\mu\nu}{\cal G}^{\mu\nu}=-4\,\,\hat{\vec l}\cdot\hat{\vec \omega}\,\,,
\ee
where $\hat{\vec l}$ and $\hat{\vec \omega}$ was defined in Eq. (\ref{19}), we observe that
\be\label{31}
\partial_\mu j^\mu = -2\,\, \hat{\vec l}\cdot\hat{\vec \omega}.
\ee
But, from the Eq. (\ref{22}) and using Eq. (\ref{08}), we can obtain, after some calculations, in the form of an Ohm's law, that
\be\label{32}
\left[ \vec l + {e\over m} \vec E \right] + {\vec u}\times \left[ \vec\omega +{e\over m}\vec B \right] =-\vec k
\ee
or
\be\label{33}
\hat{\vec l} +{\vec u} \times \hat{\vec\omega} = -{1\over \rho}\nabla \sigma -T{\nabla} s\,\,.
\ee
Thus we can see that
\be\label{34}
\partial_\mu j^\mu = -2\,\, \hat{\vec l}\cdot\hat{\vec \omega}=  \hat{\vec\omega}\cdot (\rho^{-1}\nabla\sigma) +\hat{\vec\omega} \cdot (T\nabla s)\,\,,
\ee
namely, we only have a conserved current for special vorticity fields or when we do not have contributions due to viscosity and statistical features, represented by the term $\vec k$.

So, in this case the rate of variation  of helicity to the fluid $\hat{h} =\int \hat{\vec u} \cdot \hat{\vec\omega}\,\, d^3 x$, is given by
\be\label{35}
{{d\hat{h}}\over{dt}}=-\int_{\gamma}\hat{\vec\omega} \cdot \vec k\,\,d^3x\,\,,
\ee
where $\gamma$ is the vortex tube moving with the fluid such that $\hat{\vec\omega}\cdot\hat{n}=0$ for all normal vectors to the boundary $\partial_\gamma$.  

When the fluid is incompressible with density $\rho_0$, we have 
\be\label{36}
{{d\hat{h}}\over{dt}}=-\nu\int\hat{\vec\omega}\cdot (\nabla \times \hat{\vec\omega})\,\,d^3x +\int\hat{\vec\omega} \cdot (T\nabla s)\,\,d^3x\,\,,
\ee
where $\nu =\mu /\rho_0$ is the (kinematic) viscosity. Note that the first term of the right hand side of Eq. \eqref{36} is exactly the same result found in the literature and we have an extra term due to statistical features of the fluid, the second term on the right hand side. This result has not been derived before.

\subsection{The Circulation}

The scalar functional of considerable importance in the description of vortex flows is the circulation $\cal C$ around a simple curve $\gamma$, defined as the line integral of the velocity
\be\label{37}
{\cal C}= \oint_{\gamma} \vec{v}\cdot d{\vec x}\,\,,
\ee
which is important concerning its conservation principles (Kelvin's circulation theorem). The circulation, associated with a physical quantity, calculated along the loop $\gamma$, may be zero or finite depending on whether this physical quantity is an exact differential or not. For example, if this physical quantity is $Tds$ ($T=$ temperature; $s=$ entropy), the circulation is generally finite and measures the heat gained in a quasistatic thermodynamic cycle.

We now calculate the rate of change of this circulation as the curve moves with the fluid
\be\label{38}
{d{\cal C}\over{dt}}={d\over{dt}}\oint_{\gamma} \vec{v}\cdot d{\vec x}=\oint_{\gamma} \[{{\partial\hat{\vec v}}\over{\partial t}} +(\nabla\times\hat{\vec v})\times\vec{v}\]\cdot d{\vec x}\,\,,
\ee
where the generalized velocity $\hat{\vec v}$ is expressed by
\be\label{39}
\hat{\vec v} =\vec v +{{\epsilon}\over{m}}\vec A\,\,.
\ee
Now, using that $\vec E =-{{\partial\vec A}\over{\partial t}}-\nabla\Phi$, $\vec B =\nabla\times{\vec A}$ and Eq. (\ref{21}) we obtain that
\be\label{40}
{{\partial {\hat{\vec v}}_{\alpha}}\over{\partial t}}+(\nabla\times{\hat{\vec v}}_{\alpha})\times \vec{v}_{\alpha}=-\nabla {\cal E} +\vec k_{\alpha}\,\,,
\ee
where $\cal E$ is the effective energy
\be\label{41}
{\cal E} = h +{{\epsilon}\over{m}}\Phi +{1\over 2}{\vec v}^2 \,\,.
\ee
Substituting Eq. (\ref{40}) into Eq. (\ref{38}) we have that the rate of variation of the circulation ${\cal C}$ is
\be\label{42}
{d{\cal C}\over{dt}}= -\oint_{\gamma} \nabla{\cal E}\cdot d{\vec x} +\oint_{\gamma} \vec k_{\alpha}\cdot d{\vec x} = -\oint_{\gamma} \nabla{\cal E}\cdot d{\vec x} +\int_{S} \nabla\times\vec k_{\alpha}\cdot d{\vec S}\,\,,
\ee
where 
\be\label{43}
\nabla \times \vec k_{\alpha} =\nabla T \times \nabla s + \nabla\times (\rho^{-1}\nabla \sigma) ={\vec \Gamma}_B + {\vec \Gamma}_{\nu}
\ee
where $\vec \Gamma_B= \nabla T \times \nabla s$, is the standard Biermann (baroclinic) mechanism \cite{biermann} and $\vec\Gamma_\nu = \nabla\times (\rho^{-1}\nabla \sigma)$, which was explicitly displayed on the right hand side, are the possible sources of the circulation $\cal C$. So, we can write that
\be\label{44}
{d{\cal C}\over{dt}}= -\oint_{\gamma} \nabla{\cal E}\cdot d{\vec x} +\int_{\partial\gamma} \vec \Gamma_B\cdot d^2{\vec x} + \int_{\partial\gamma} \vec \Gamma_{\nu}\cdot d^2{\vec x}\,\, ,
\ee
where $\partial_\gamma$ is the surface that moves along the fluid. 

We can observe from Eq. \eqref{44} that the first term on the right hand side contributes to the rate of change of the circulation if $\oint_{\gamma} \nabla{\cal E}\cdot d{\vec x}=\int_{\gamma}\,d\,{\cal E}\neq 0$, that is, if the fluid-dynamic force derived from the energy density $\cal E$ is not an exact differential.

Besides, we have two other contributions to the rate of variation of circulation, the second term on the right side, which is a form of source that could, in principle, generate vorticity. It represents the flow of $\vec \Gamma_B$ through the surface $\partial_\gamma$. This term is zero when it is a barotropic fluid, where the density only depends on the pressure and the temperature is constant. 

The last term of the right hand side of Eq. \eqref{44} is also a source term for a contribution due to the viscosity of the fluid, and represents a flow of $\vec \Gamma_{\nu}$ through the surface $\partial\Gamma$. Notice that, by considering for simplicity the case of an ideal fluid with density $\rho_0$, we have the term $\vec \Gamma_{\nu}\approx\nabla^2\hat{\vec\omega}$. In this case, at the beginning, when the vorticity is still zero $\vec \Gamma_{\nu}$ is necessarily negligible. However, being proportional to the generated vorticity, the dissipative term will become progressively large as the fluid builds up vorticity.

\section{Final discussions}\label{s3}

In this paper we have obtained the analytical form of the source terms which is a new result that helps in the calculation of the generalized fields fo the fluid.   After that, we have constructed the Navier-Stokes equation for the charged fluid immersed in an electromagnetic field.  And we have calculated the wave equations for the velocity and vorticity fields.   The gauge choice dynamics were analyzed.

Finally, we have analyzed the time evolution of helicity and circulation from Maxwell's type formulation for the equations of a compressive fluid, charged in an interaction with an electromagnetic field. We can see that for both helicity and circulation there are terms that, in principle, can be considered as source terms or creation of circulation in the dynamics of fluids.

\section{ Acknowledgments}

\ni E.M.C. Abreu thanks CNPq
(Conselho Nacional de Desenvolvimento Cient\' ifico e
Tecnol\'ogico), Brazilian scientific support federal agency, for
partial financial support, Grants numbers 302155/2015-5
and 442369/2014-0, and  the
hospitality of Theoretical Physics Department at Federal
University of Rio de Janeiro (UFRJ), where part of this work was
carried out.


\begin{thebibliography} {99}

\bibitem{Kambe}    T. Kambe, Fluid Dyn. Res. {\bf 42}, 055502 (2010).

\bibitem{Thompson}    R. J. Thompson and T. M. Moeller, Phys. of Plasmas {\bf 19}, 010702 (2012); {\bf 19}, 082116 (2012).

\bibitem{Marmanis}   H . Marmanis, Phys. Fluids. {\bf 10}, 1428 (1998).

\bibitem{jnpp}   R. Jackiw, V. P. Nair, S. Y. Pi, and A. P. Plolychronakus, J. Phys. A {\bf 37}, R327 (2004).

\bibitem{lighthill}   M. J. Lighthill, Proc. R. Soc. A {\bf 211}, 564 (1952); {\bf 222}, 1 (1954).

\bibitem{hawking}   S. W. Hawking, Phys. Rev. D {\bf 13}, 191 (1976).

\bibitem{beckenstein}   J. D. Bekenstein, Phys. Rev. D {\bf 9}, 3292 (1974).

\bibitem{beckenstein2}   J. D. Bekenstein, Phys. Rev. D {\bf 23}, 287 (1981).

\bibitem{fh}   J. A. de Freitas Pacheco and J. E. Horvath, Class. Quant. Grav. {\bf 24}, 5427 (2007).

\bibitem{gs}   P. F. Gonz\'alez-Dias and C. L. Sig\"uenza, Nucl. Phys. B {\bf 697}, 363 (2004).


\bibitem{albert1}   E. M. C. Abreu, J. A. Neto, A. C. R. Mendes and N. Sasaki, Phys. Rev. D {\bf 91}, 125011 (2015).

\bibitem{mahajan}   S. M. Mahajan, Phys. Rev. Lett. 90 (2003) 035001.

\bibitem{Albert1}   A. C. R. Mendes, C. Neves. W. Oliveira and F. I. Takakura, Braz. J. Phys. {\bf 33}, 346 (2003).


\bibitem{albert2} A. C. R. Mendes, F. I. Takakura, E. M. C. Abreu and J. A. Neto, Eur. Phys. Lett. {\bf 116}, 20004 (2016).

\bibitem{albert3} A. C. R. Mendes, F. I. Takakura, E. M. C. Abreu and J. A. Neto, Eur. Phys. Lett. {\bf 380}, 12 (2017).

\bibitem{Landau}   L. D. Landau and E. M. Lifshits, {\it Fluid Mechanics}, Pergamon Press, Oxford, 1980.


\bibitem{biermann}    L. Biermann,  Z. Naturforsch. Teil A, {\bf 5}, 65 (1950).

\end{thebibliography}
\end{document}